# Improved method for phase wraps reduction in profilometry

Guangliang Du[a], Minmin Wang[a], Canlin Zhou [a*],Shuchun Si[a], Hui Li[a], Zhenkun Lei[*b],Yanjie Li[c]

[a] School of Physics, Shandong University, Jinan 250100, China

[b] Department of Engineering Mechanics, Dalian University of Technology, Dalian 116024, China

[c] School of Civil Engineering and Architecture, University of Jinan, Jinan, 250022, China

[*]Corresponding author: Tel: +8613256153609; E-mail address: canlinzhou@sdu.edu.cn；leizk@163.com

## Abstract

In order to completely eliminate, or greatly reduce the number of phase wraps in 2D wrapped phase map, Gdeisat et al. proposed an algorithm, which uses shifting the spectrum towards the origin. But the spectrum can be shifted only by an integer number, meaning that the phase wraps reduction is often not optimal. In addition, Gdeisat's method will take much time to make the Fourier transform, inverse Fourier transform, select and shift the spectral components. In view of the above problems, we proposed an improved method for phase wraps elimination or reduction. First, the wrapped phase map is padded with zeros, the carrier frequency of the projected fringe is determined by high resolution, which can be used as the moving distance of the spectrum. And then realize frequency shift in spatial domain. So it not only can enable the spectrum to be shifted by a rational number when the carrier frequency is not an integer number, but also reduce the execution time. Finally, the experimental results demonstrated that the proposed method is feasible.

## Keywords

phase unwrapping; zero-padding; Fourier transform; carrier-frequency; profilometry

## 1. Introduction

Phase-based fringe projection technique is an important method in three-dimensional (3D) shape measurement. It has been extensively investigated and widely used in aerospace, biological engineering, military reconnaissance and other fields because of its full field, real-time, non-contact, simple device and higher accuracy [1-4]. The basic procedure is as follows: first, the sinusoidal fringe is projected onto the surface of the tested object, and the deformed fringe is recorded by a camera, and then the phase information of the deformed fringe is demodulated by the four-step phase shift method or the Fourier transform method. Since the phase obtained by phase demodulation method is wrapped in $(-\pi,\pi)$, phase unwrapping is necessary to obtain the continuous phase [5-7]. In recent decades, a variety of unwrapping algorithms are proposed. Most phase-unwrapping algorithms can be classified into two categories: temporal phase unwrapping and spatial phase unwrapping. The spatial phase unwrapping is a process of integral accumulation, once one of these points appears error, this error will spread to the following points, which will affect the calculation of the phase, and it may lead to the phenomenon of wire drawing [8-9]. In the actual measurement, discontinuous morphology, noise and shadow may lead to errors. To avoid the impact of these factors on the unwrapping process, some scholars proposed local spatial phase unwrapping and global spatial phase unwrapping method. The local spatial phase unwrapping method includes branch-cut method [10], quality guided method [11] and so on. The global spatial phase unwrapping method mainly includes the least square method [12] and the minimum norm method [13]. Although these methods have a certain improvement, there are some limitations. Saldner et al. [14] proposed the temporal phase unwrapping method, which is made in the temporal domain, a sequence of maps is acquired while the fringe pitch is changed. Then the phase at each pixel is unwrapped along the time axis to avoid the spread of errors. However, the method needs multiple frames of fringe images which would take much time. Although there are many phase unwrapping algorithms, no algorithm can be applied to address all problems. If the number of phase wraps in the phase diagram can be eliminated or reduced, it will undoubtedly be beneficial to reduce the unwrapping time and improve the noise performance. Gdeisat et al. [15] proposed phase wraps reduction method to completely eliminate, or greatly reduce the number of phase wraps, which would make the phase unwrapping simple and fast. In some cases, all of the phase wraps are eliminated and there is therefore no need to unwrap the resultant phase map. However, according to our own experience with the method, Gdeisat's method has the following disadvantages: (1) This method needs Fourier transform, inverse Fourier transform, select and shift the spectral components, these procedures increase the calculation complexity and the processing time consuming as well. (2) The spectrum can be shifted only by an integer number in Gdeisat's method, meaning that the phase wraps reduction is often not optimal.

Here, in order to improve the calculation efficiency as well as simplify its procedures, we present an improved method for phase wraps elimination or reduction. This method is a good solution to the problems of Gdeisat's method. The capability of the presented method is demonstrated by both theoretical analysis and experiments.

The paper is organized as follows. Section 2 introduces the principle of the system. Section 3 presents the experimental results. Section 4 summarizes this paper.

# 2. Theory

## 2.1 Gdeisat's method

Gdeisat et al. proposed an algorithm to completely eliminate, or greatly reduce the number of phase wraps in the phase diagram using spectrum-shifting, the basic process is as follows, first, convert the wrapped phase map into the complex array, and calculate the 2D Fourier transform. Then shift the spectrum towards the origin and make the inverse Fourier transform. Finally, extract the phase using the arctangent function. After this, the number of phase wraps in the new phase map is greatly reduced or eliminated, which would make the phase unwrapping simple, fast, and even without phase unwrapping. In [15], when the maximum phase change in the deformed fringe patterns is less than $\pi$ radians, by applying Gdeisat's method to the wrapped phase map, all the phase wraps in the image are eliminated. The resultant phase map therefore does not require the application of any phase unwrapping algorithms. We think the condition is same with the initial phase in [4][16].

Fig.1 shows the optical path of phase measuring profilometry, where P is the projection center of the projector, C is the camera imaging center, and D is an arbitrary point on the tested object. Its modulation phase can be calculated by,

$$\phi_{BD} = \frac{dh}{(l-h)l_0}\phi_0 \tag{1}$$

where $l_0$ is the maximum width of field-of-view (FOV) observed by CCD camera, $\phi_0$ is the corresponding phase values within $l_0$. If $\phi_{BD} < 2\pi$ ,the resultant phase map does not require any phase unwrapping after applying Gdeisat's method. On the contrary, if $\phi_{BD} > 2\pi$, Gdeisat's method can greatly reduce the number of phase wraps, so the speed and accuracy are greatly improved when the resultant phase map is processed using the unwrapping algorithm.

## 2.2 Our method

In [15], Gdeisat's method can eliminate, or greatly reduce the number of phase wraps in the phase map, but it also has some disadvantages, that is, (1) this method needs twice Fourier transform, and it needs to search and select the spectral components, which will occupy large amount of processing time. (2) because when the computer processes the digital images, it uses the discrete Fourier transform, the spectrum can be shifted only by an integer number in Gdeisat's method. But the actual frequency may be a fraction, that is, it is a rational number. So when we use Gdeisat's method, the phase wraps reduction is often not optimal. Based on our analysis of the ref.[15], we put forward the corresponding solutions.

Assuming that the four phase-shifted fringe images is as follows [15],

$$\begin{aligned}
I_1(x,y) &= \cos[2\pi f_x x + 2\pi f_y y + \beta\varphi(x,y)] \\
I_2(x,y) &= \cos[2\pi f_x x + 2\pi f_y y + \beta\varphi(x,y) + \frac{\pi}{2}] \\
I_3(x,y) &= \cos[2\pi f_x x + 2\pi f_y y + \beta\varphi(x,y) + \pi] \\
I_4(x,y) &= \cos[2\pi f_x x + 2\pi f_y y + \beta\varphi(x,y) + \frac{3\pi}{2}]
\end{aligned} \tag{2}$$

Where $f_x$ and $f_y$ are the spatial carrier frequency along the X axis and the Y axis respectively, $\beta$ is the modulation index. The wrapped phase can be extracted using the well-known Eq.(3) below by the four-step phase shift algorithm.

$$\varphi_w(x,y) = \tan^{-1}[\frac{I_4 - I_2}{I_3 - I_1}] = W(2\pi f_x x + 2\pi f_y y + \beta\varphi(x,y)) \tag{3}$$

Where $\tan^{-1}$ is the four quadrant arctangent operator, and $\varphi_w(x,y)$ is the wrapped phase which is wrapped in $(-\pi,\pi)$ . The following three equations convert the wrapped phase map into the complex array $\varphi_{wc}(x,y)$ .

$$
\begin{aligned}
&R(x,y)=\cos[\varphi_w(x,y)]\\
&I(x,y)=\sin[\varphi_w(x,y)] \qquad (4)\\
&\varphi_{wc}(x,y)=R(x,y)+jI(x,y)
\end{aligned}
$$

Where j is equal to $\sqrt{-1}$.

In Gdeisat's method, first, make the Fourier transform to $\varphi_{wc}(x,y)$ as shown in Eq.(5).

$$\Phi_{wc}(u,v)=\xi[\varphi_{wc}(x,y)] \qquad (5)$$

where $\xi$ [.] is the 2D Fourier transform operator, and the terms u and v are the vertical and horizontal frequencies respectively. The 2D Fourier transform of the wrapped phase $\Phi_{wc}(u,v)$ is shifted towards the origin using the indices $u_0$ and $v_0$, then do the inverse Fourier transform.

$$\varphi_{wcs}(x,y)=\xi^{-1}[\Phi_{wc}(u+u_0,v+v_0)] \qquad (6)$$

where $\xi^{-1}$[.] is the inverse 2D Fourier transform operator, $u_0$, $v_0$ is moving distance. We can obtain the phase map as follows.

$$\varphi_{ws}(x,y)=\tan^{-1}\frac{I\{\varphi_{wcs}(x,y)\}}{R\{\varphi_{wcs}(x,y)\}}=W(2\pi(f_x-u_0)x+2\pi(f_y-v_0)y+\beta\varphi(x,y)) \qquad (7)$$

where I{.} represents the imaginary part, and R{.} represents the real part of the complex array $\varphi_{wcs}(x,y)$.

From Eq.(7), we can see, shifting the spectrum in the frequency domain towards the origin equals to decreasing the spatial carrier frequency of projected fringe, thus the number of phase wraps in the phase map is reduced.

Through the analysis, it is easy to see that the spectrum can be shifted only by an integer number with the discrete Fourier transform, while the actual frequency in 3D morphology experiment may not exactly be an integer, but a rational number. Thus when we use Gdeisat's method, the phase wraps reduction is often not optimal. Besides, Gdeisat's method requires Fourier transform, spectrum search, shifting and the inverse Fourier transform. Therefore, in the actual operation, the process is relatively complex. To solve this problem, we simplify and speed up the implementation process in spatial domain by the frequency shift property of 2D Fourier transform[17]. The frequency shift property of 2D Fourier transform can be written as:

$$F(u+u_0,v+v_0)=\xi[f(x,y)e^{(-j2\pi(u_o x/m+v_0 y/n))}] \qquad (8)$$

where $F(u,v)$ is the Fourier transform of $f(x,y)$, m, n are the length of $f(x,y)$ along the X axis and the Y axis respectively.

By Eq.(8), we can directly obtain $\varphi_{wcs}(x,y)$ by multiplying $e^{(-j2\pi(u_o x/m+v_0 y/n))}$ by the result $\varphi_{wc}(x,y)$ of Eq.(4).

$$\begin{aligned}\Phi_{wc}(u+u_0,v+v_0)&=\xi[\varphi_{wc}(x,y)e^{(-j2\pi(u_o x/m+v_0 y/n))}]\\ \varphi_{wcs}(x,y)&=\xi^{-1}[\Phi_{wc}(u+u_0,v+v_0)]=\varphi_{wc}(x,y)e^{(-j2\pi(u_o x/m+v_0 y/n))}\end{aligned} \tag{9}$$

By Eq.(9), we can realize frequency shift in spatial domain by the frequency shift property of 2D Fourier transform. It is obvious that our method does not require Fourier transformation, spectrum selection, spectrum shift and inverse Fourier transformation, and $u_0$, $v_0$ can be a rational number, so the proposed method overcomes the limitation of the Gdeisat's method on the frequency shift value.

In order to determine the actual frequency of the deformed fringe pattern in 3D profilometry experiment, Fan et al.[18] proposed a spectrum centroid method for estimating the carrier frequency, still needed Fourier transform. Although the accuracy is high, the computational burden is large. Feng et al.[19] proposed an accurate method to estimate the carrier frequency based on zero padding[20]. According to[19], pad $\varphi_w(x,y)$ in Eq.(3) with zeros,

$$\varphi_{wz}(x,y)=\begin{cases}\varphi_w(x,y); 0\le x\le M-1, 0\le y\le N-1\\ 0; M\le x\le kM-1, N\le y\le kN-1\end{cases} \tag{10}$$

Where M, N is the value of $\varphi_w(x,y)$ along the X axis and the Y axis respectively, k is an integer.

Make the Fourier transform to $\varphi_{wz}(x,y)$,

$$\Phi_{wz}\left(\frac{u}{kM},\frac{v}{kN}\right)=\sum_{m=0}^{kM-1}\sum_{n=0}^{kN-1}\varphi_{wz}(x,y)\exp\left[-j2\pi\left(\frac{u}{kM}x+\frac{v}{kN}y\right)\right] \tag{11}$$

From Eq.(11), we can see that zero padding in the spatial domain will achieve up-sampling in the frequency domain[19-20]. In [19], taking k=10, the accuracy is improved by 10 times. However, if we use the method in [19], the fringe pattern will become very large after zero padding, and when we make the Fourier transform to the padded fringe pattern, computation burden will be quite large. So Feng's method requires high performance computer hardware, and the speed is very slow.

Therefore, in order to determine the actual carrier frequency of the deformed fringe pattern, we improve the zero padding method in the reference[19]. The basic idea is, the two-dimensional Fourier transform is replaced by 2 one-dimensional (1D) Fourier transform, and the horizontal and vertical components of the fringe carrier frequency are estimated by 1D Fourier transform respectively. The specific procedure is as follows,

(1) take the middle of the row and the column of $\varphi_w(x,y)$ in Eq.(3), pad them with zeros respectively, thus obtain 2 1D vector data.

(2) make 1D Fourier transform respectively to the 2 1D vector data in step (1), estimate

their spectrum peak position, and obtain $u_0$, $v_0$.

As a result of only 2 1D Fourier transform, even if taking k=100, zero padding is still very simple, and the calculation speed is still very fast.

The following experiment is used to verify the proposed algorithm.

## 3. Experiments

In this section, for evaluating the real performance of our method, we test our method on a series of experiments. Below, we will describe these experiments and practical suggestions for the above procedure.

We develop a fringe projection measurement system, which consists of a DLP projector (Optoma EX762) driven by a computer and a CCD camera ( DH-SV401FM). Fig. 1 shows the schematic of fringe-projection profilometry system. In our experiments, d=60cm, l=70cm, l0=20cm, and the captured image is 688 pixels wide by 582 pixels high. The surface measurement software is programmed by Matlab with I5-4570 CPU @ 3.20 GHz.

First, we do an experiment on a face model with smooth shapes, the biggest height is about 3cm, the frequency of projected fringe $\phi_0$ =12π. By calculation, $\phi_{BD}$ in Eq.(1) is less than 2π. In theory, the resultant phase map does not require any phase unwrapping after applying Gdeisat's method. Fig. 2 shows the captured image. The wrapped phase map by four-step phase shift method is shown in Fig. 3.

Take the middle of the row(the $300^{th}$ row) in Fig. 3, pad it with zeros(k=10), the result is shown in Fig. 4.

Fig. 5 shows the spectrum of Fig. 4 by 1D Fourier transform. We can see, the resolution of frequency domain is improved by ten times due to zero padding. By estimating the peak of the spectrum, determine $v_0$=4.7.

With the same method to take the middle of the column (the $300^{th}$ column) in Fig. 3, pad it with zeros(k=10), we can determine $u_0$=3.8.

Then we put $u_0$ and $v_0$ into Eq.(9), the phase map $\phi_{wcs}(x, y)$ is shown in Fig. 6. The phase wraps are completely eliminated.

For comparison, according to Gdeisat's method, take $u_0$=4, $v_0$=5, get the phase data as shown in Fig. 7. From Fig. 7, we can see, there are still phase wraps after applying Gdeisat's method. Because the actual carrier frequency is not an integer, errors appear in the results, and it requires phase unwrapping to obtain the continuous phase. From Fig. 6, we can find that with the proposed method, the carrier frequency can be determined more accurately, and the phase wraps are completely eliminated. So the resultant phase map no longer requires any phase unwrapping.

In order to compare the time between Gdeisat's method and the proposed method, we conduct a comparative experiment.

Table 1 lists the comparisons of time consuming between these two methods. In the comparisons, all processed fringe patterns have the pixels sizes of 688 x 582, the computational platform is a personal laptop with Intel Core i5-4570 CPU at 3.20GHz and a 4GB RAM. We use MATLAB 2014a on the same computer to process the same fringe pattern.

| method | Time consuming |
|---|---|
| Gdeisat's method | 1.390s |
| The proposed method | 0.672s |

Table.1. Comparisons of time consuming of the two methods

As can be seen from table 1, the difference of time consuming between Gdeisat's method and the proposed method is obvious, the speed of the proposed algorithm is improved by about 50%. In the proposed method, the calculation procedures such as Fourier transformation, frequency selection and inverse transformation are not required, therefore, it can save large amount of processing time.

For a more complete verification to the proposed method, we do another experiment on a face model with bigger height and complex shapes, the biggest height is about 5cm, the frequency of projected fringe $\phi_0$ =12π. By calculation, $\phi_{BD}$ in Eq.(1) is more than 2π. In theory, the resultant phase map is still wrapped after applying Gdeisat's method. Fig. 8 shows the captured image. The wrapped phase map by four-step phase shift method is shown in Fig. 9. Then we unwrap the phase map by Herráez's method[21] 10 times, the total processing time is 19.72s.

Fig. 10 shows the 300$^{th}$ row of Fig. 9 after zeros padding (k=10).

Fig. 11 is the spectrum of Fig. 10 by 1D Fourier transform. By calculating the peak of the spectrum, we can determine $v_0$=5.3. With the same method, we can determine $u_0$=4.5.

Put $u_0$ and $v_0$ into Eq.(9), the phase map $\phi_{wcs}(x, y)$ is shown in Fig. 12. We also unwrap the phase map $\phi_{wcs}(x, y)$ by Herráez's method[21] 10 times, the total processing time is 16.55s. The resulting unwrapped-phase map produced from Fig. 12 is shown in Fig. 13

For comparison, according to Gdeisat's method, take $u_0$=5, $v_0$=5, obtain the phase data as shown in Fig. 14. We also use Herráez's method[21] to unwrap it 10 times, and the total processing time is 17.67s. The unwrapped phase map produced from Fig. 14 is shown in Fig. 15.

From Fig. 12 and Fig. 14, we can see, because the tested object is too high, the phase wraps are not eliminated completely after the spectrum shift , but the phase wraps reduction is optimal by the proposed method, which will be more conducive to the processing of phase unwrapping algorithm.

# 4. Conclusion

In this paper, we propose an improved method for phase wraps elimination or reduction, which is an extension of Gdeisat's method. Our method overcomes the main disadvantages that Gdeisat's method encounters. The proposed method estimates the actual frequency of the deformed fringe pattern more accurately with zero padding method, eliminates Fourier transformation, inverse Fourier transformation and frequency selection in Gdeisat's method, and achieves the frequency shift by a simple multiply operation in spatial domain by the frequency shift property of 2D Fourier transform. So the proposed method not only improves the operation speed, but also adapts to the situation when the spectrum shift distance is a rational number, which would make the phase wraps reduction optimal.

## Acknowledgment

This work was supported by the National Natural Science Foundation of China (Grant nos. 11302082 and 11472070). The support is gratefully acknowledged.

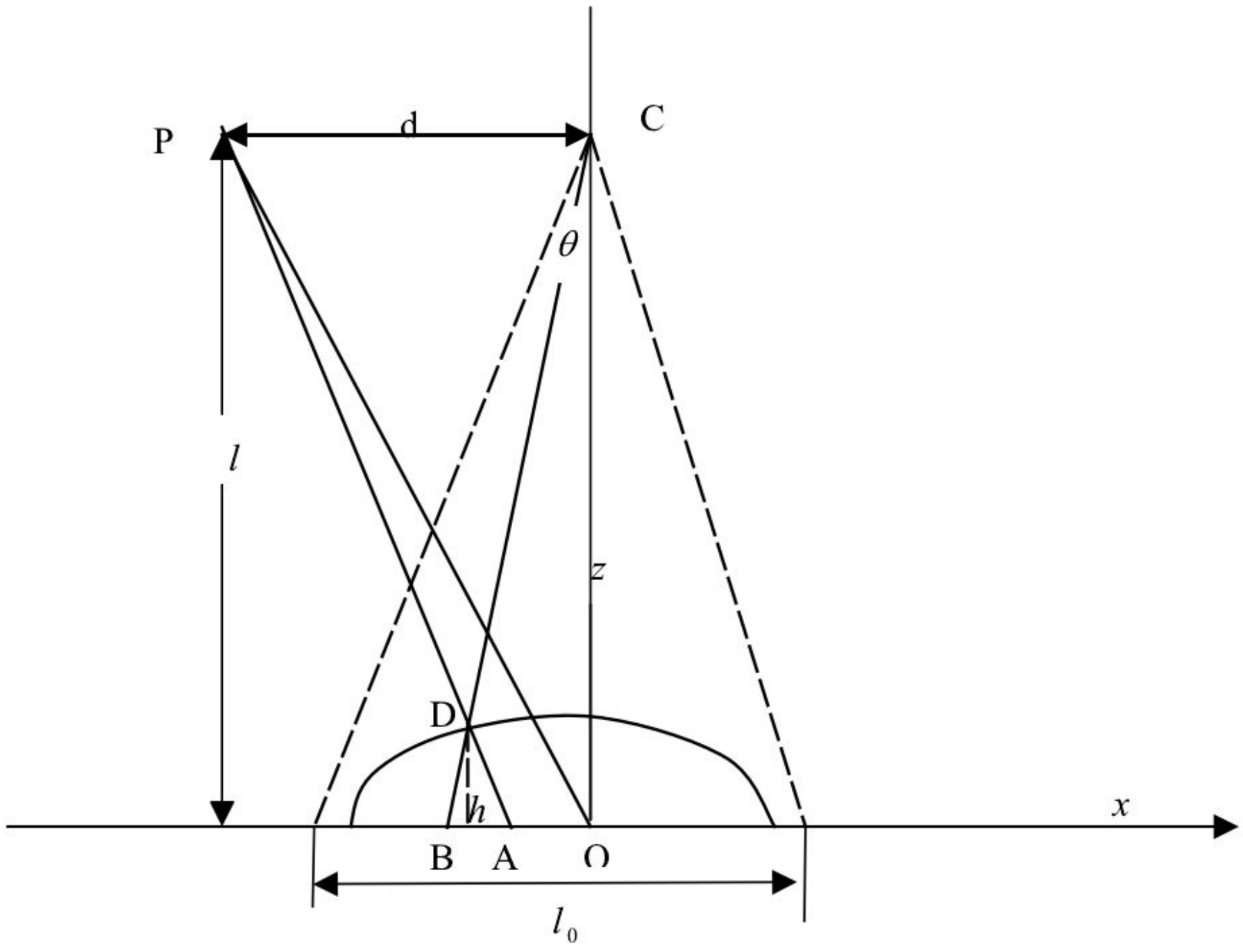

Fig.1.Optical path of phase measuring profilometry

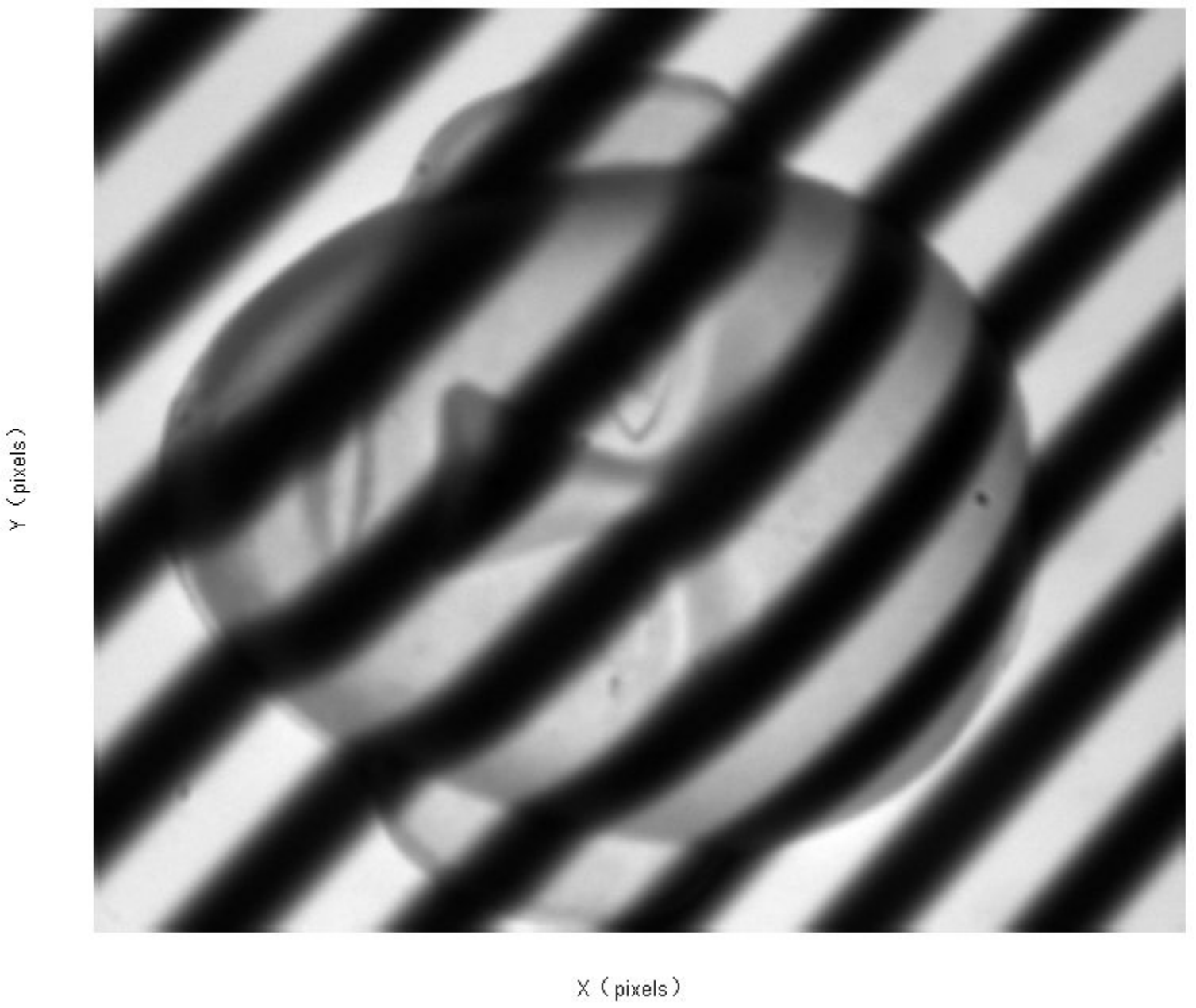

Fig. 2. The first pattern of the sinusoidal fringe

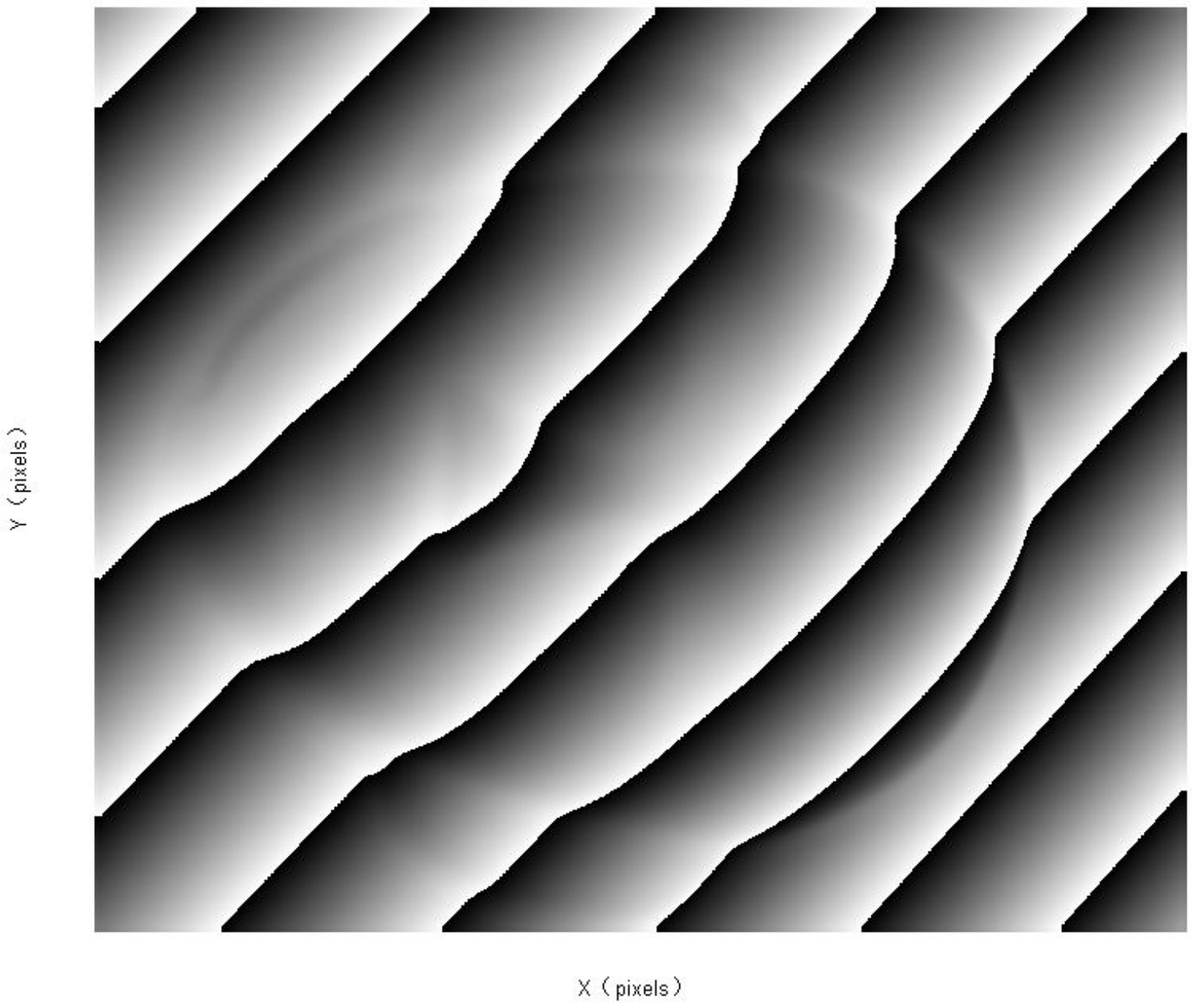

Fig. 3. The wrapped phase map

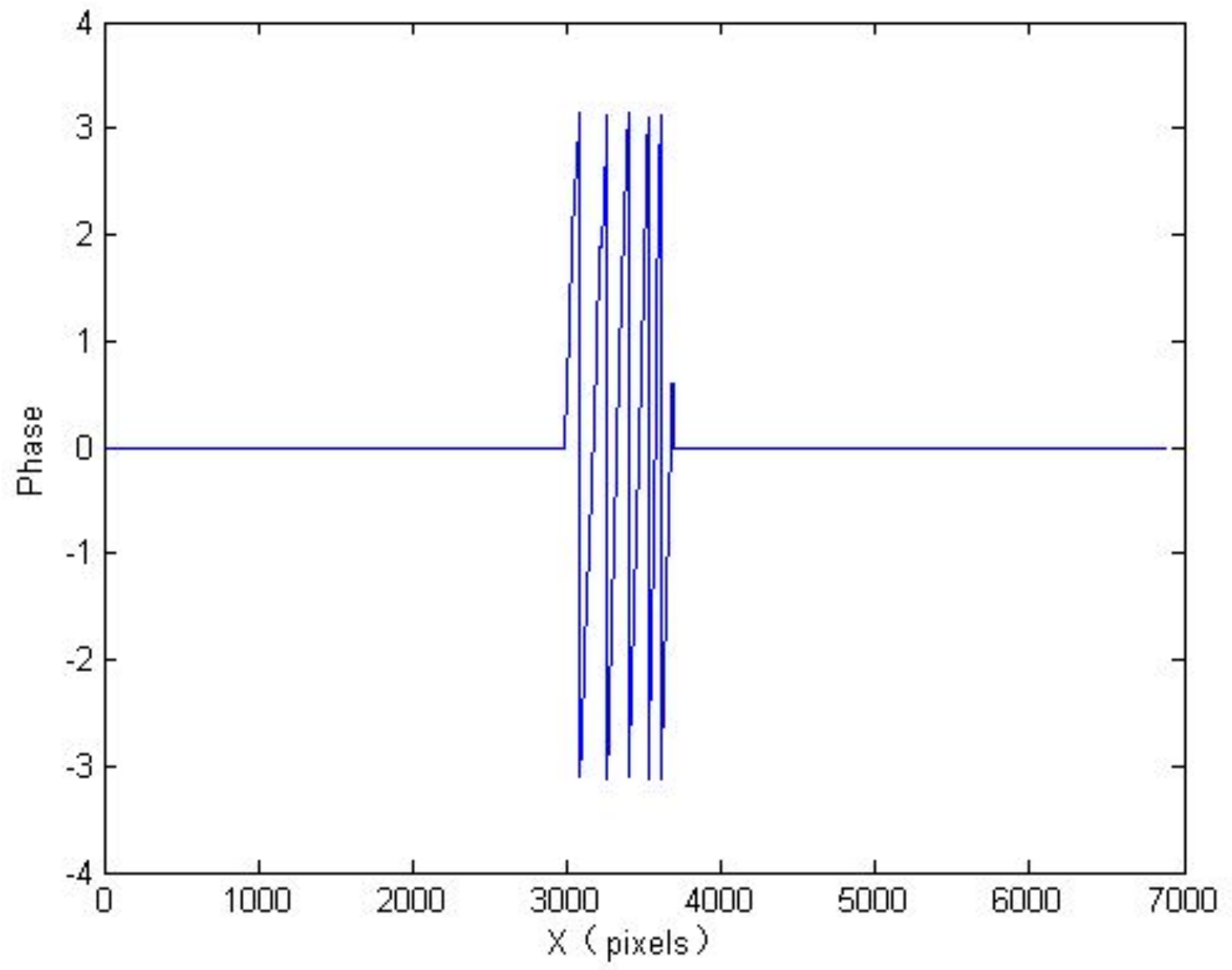

Fig. 4. The 300th row of Fig. 3 after zero padding

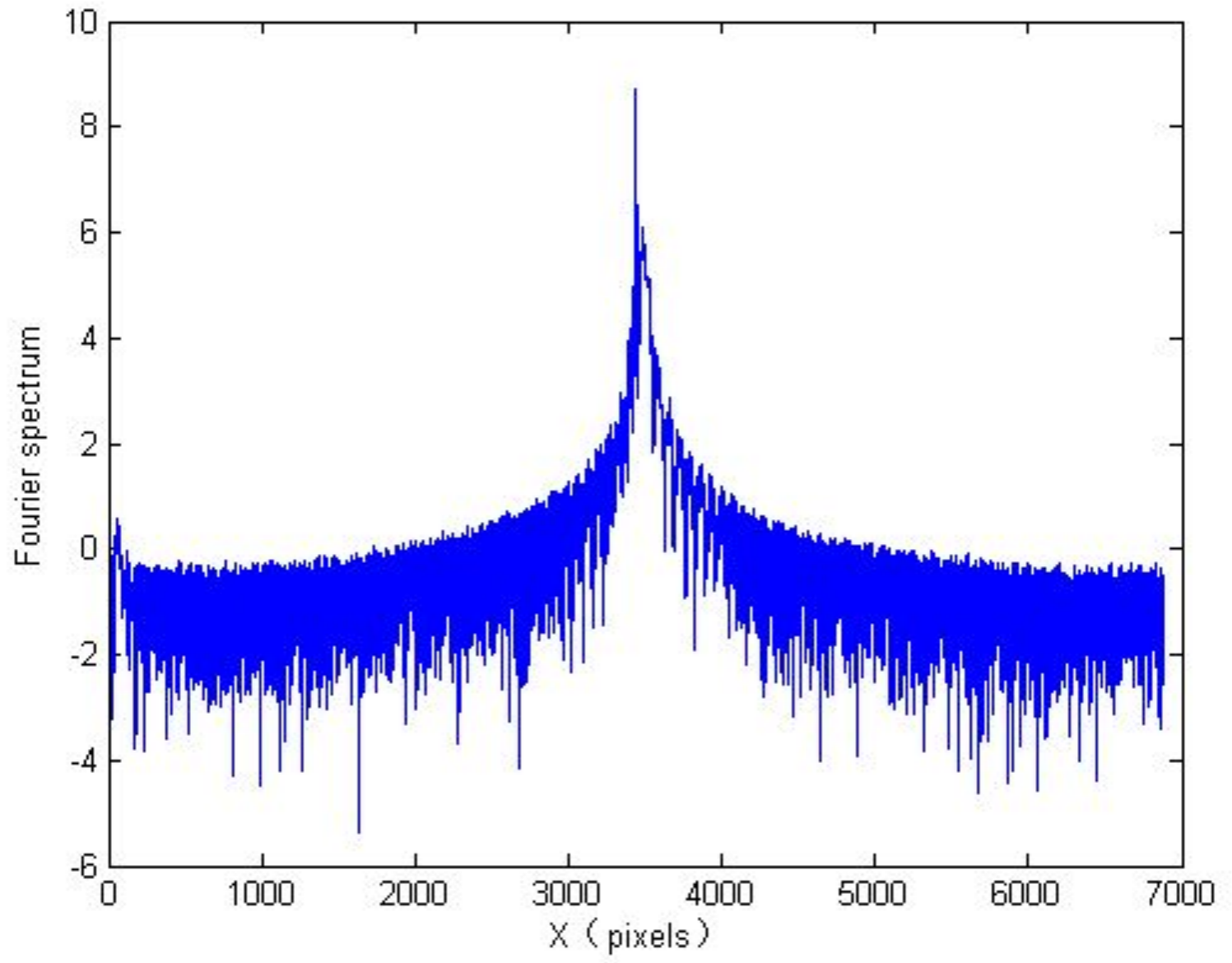

Fig.5. The spectrum of Fig. 4

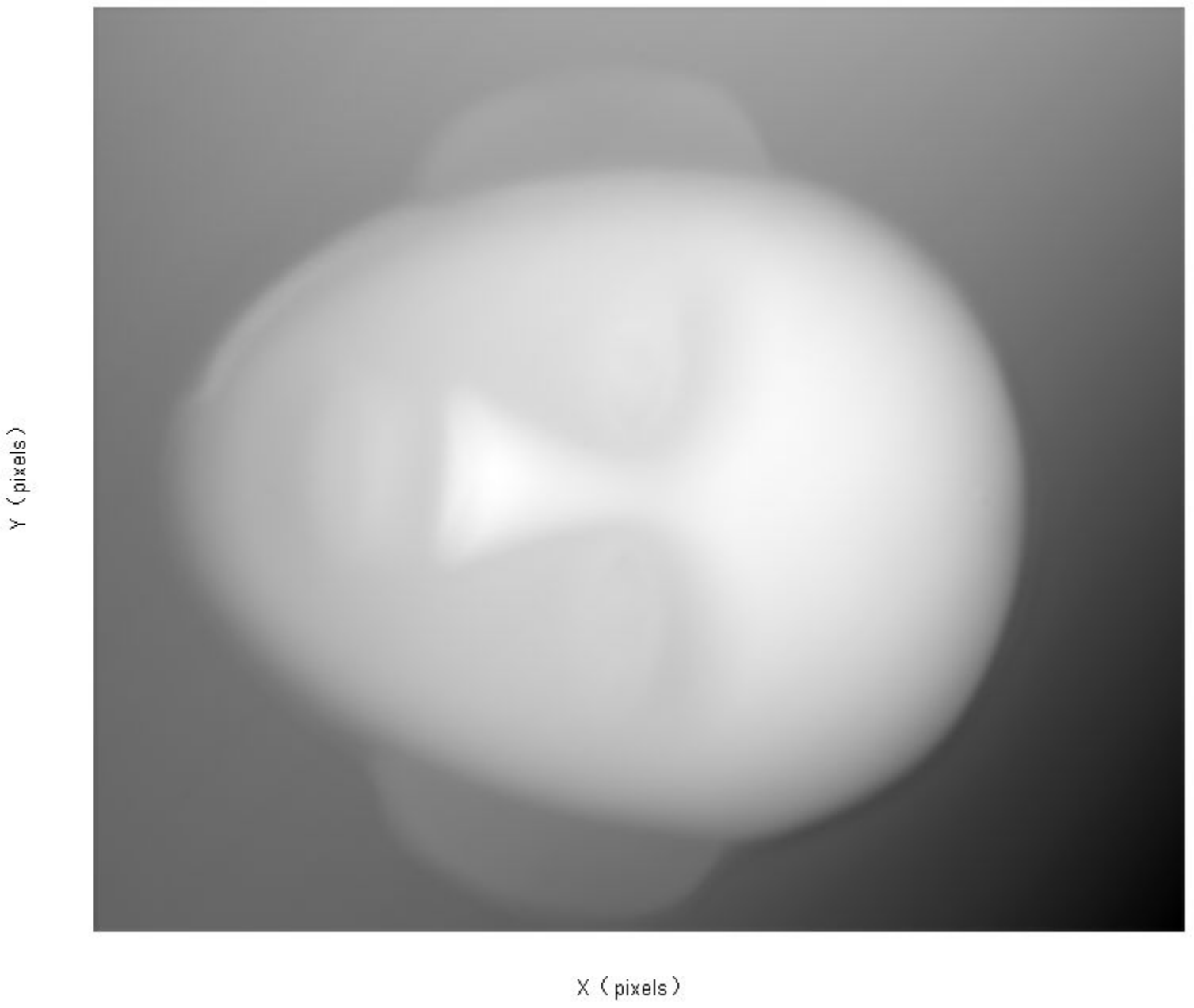

Fig.6. The phase map obtained by the proposed method

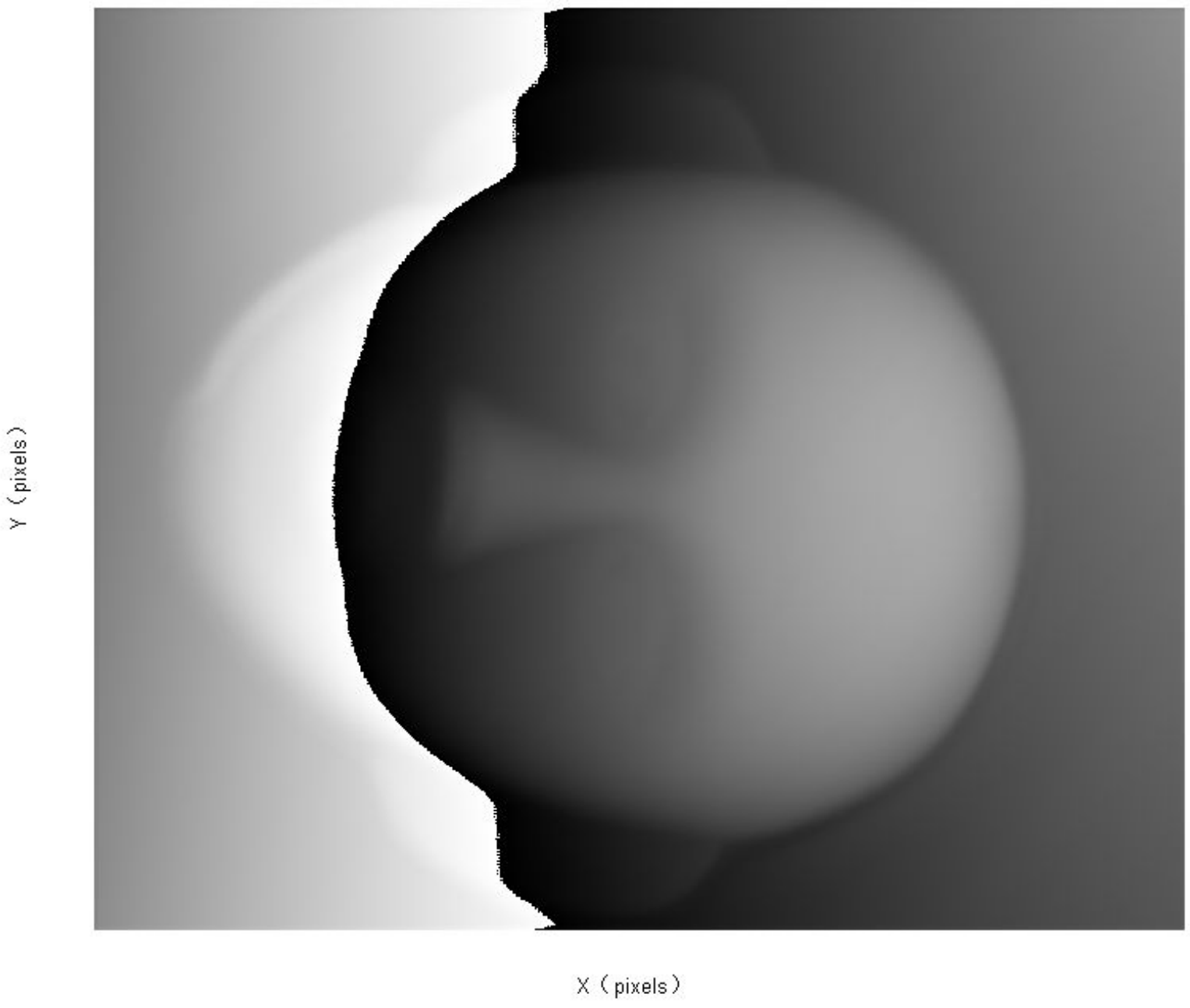

Fig. 7. The phase map obtained by Gdeisat’s method

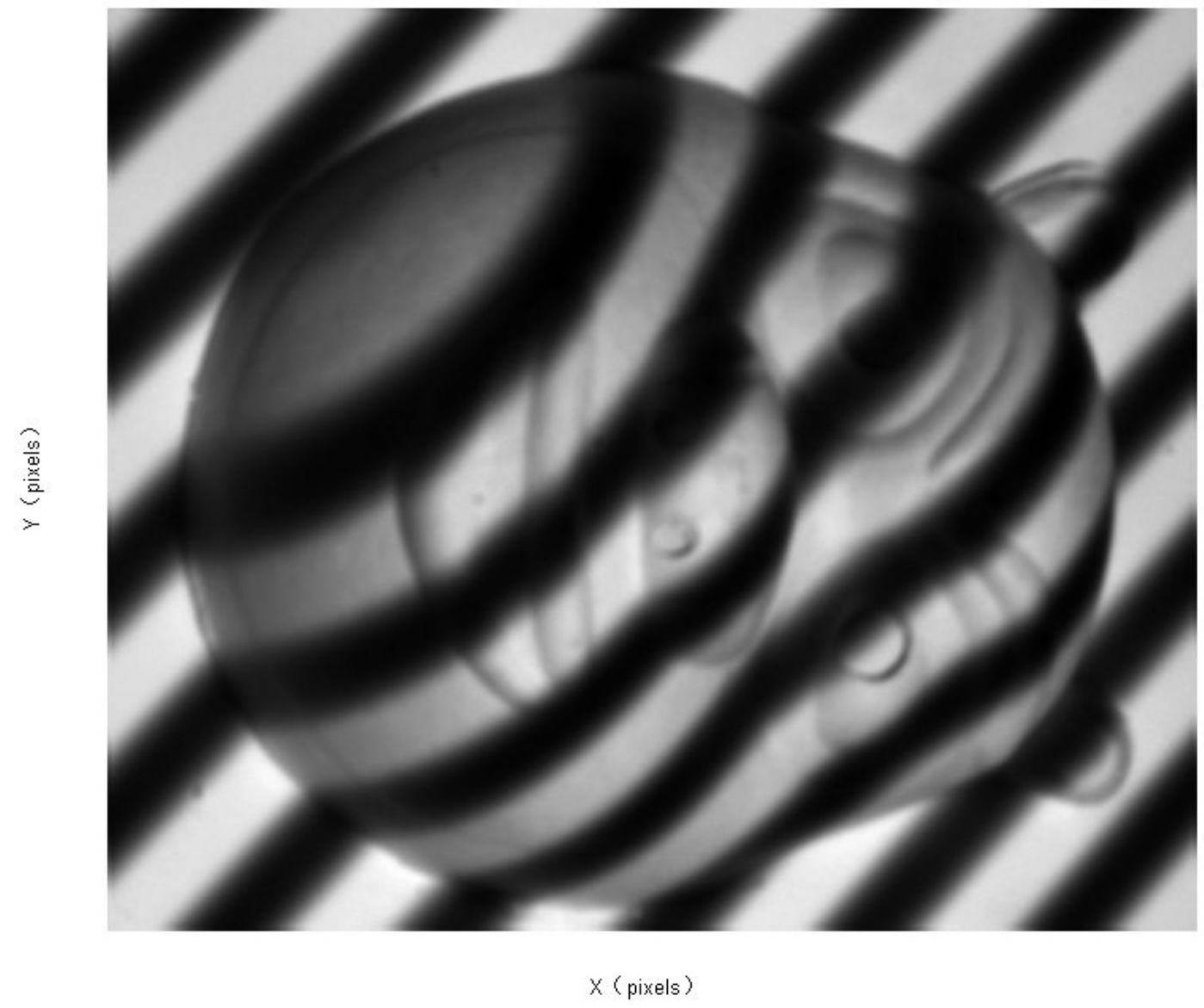

Fig.8. The first pattern of the sinusoidal fringe

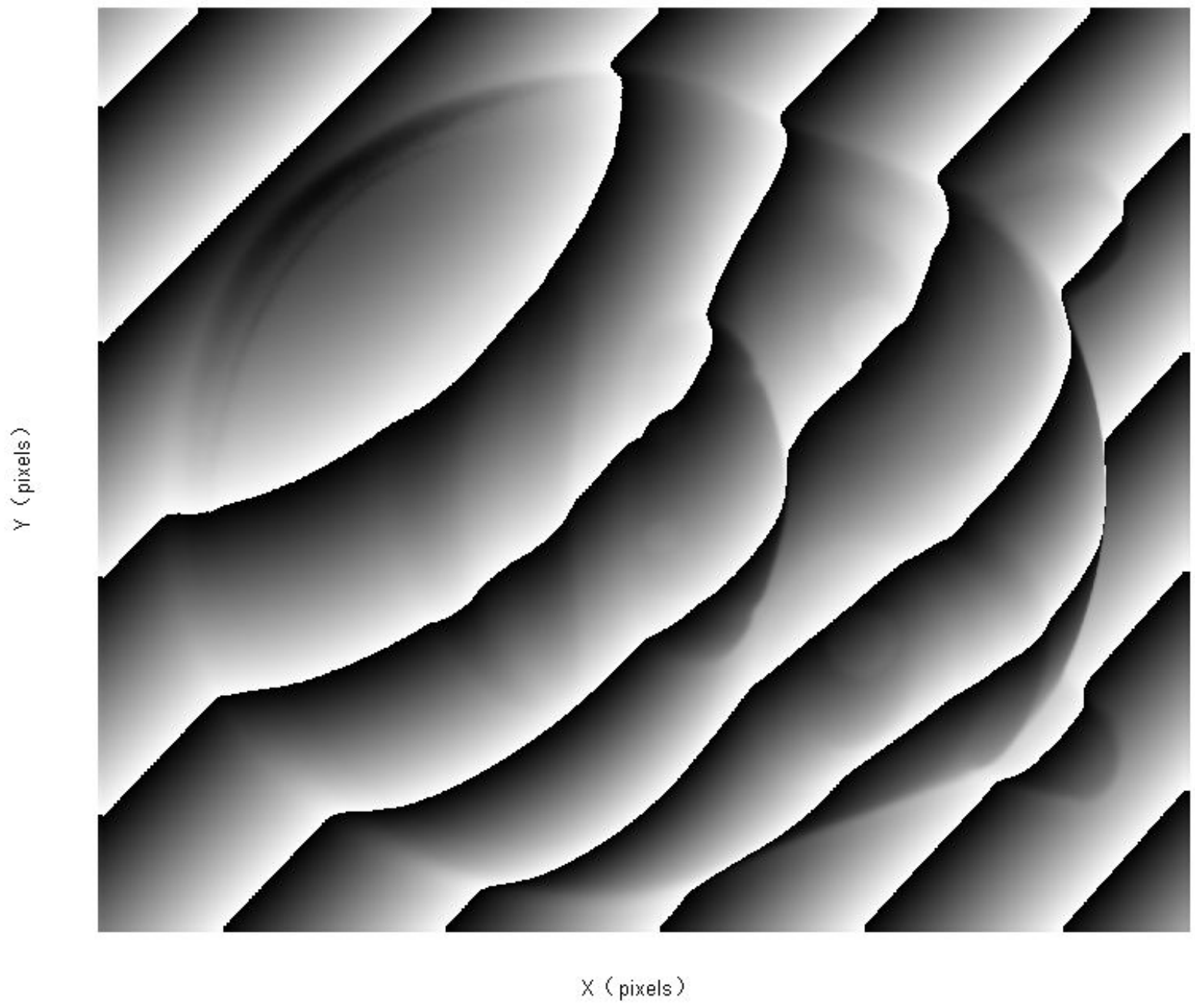

Fig.9. The wrapped phase map

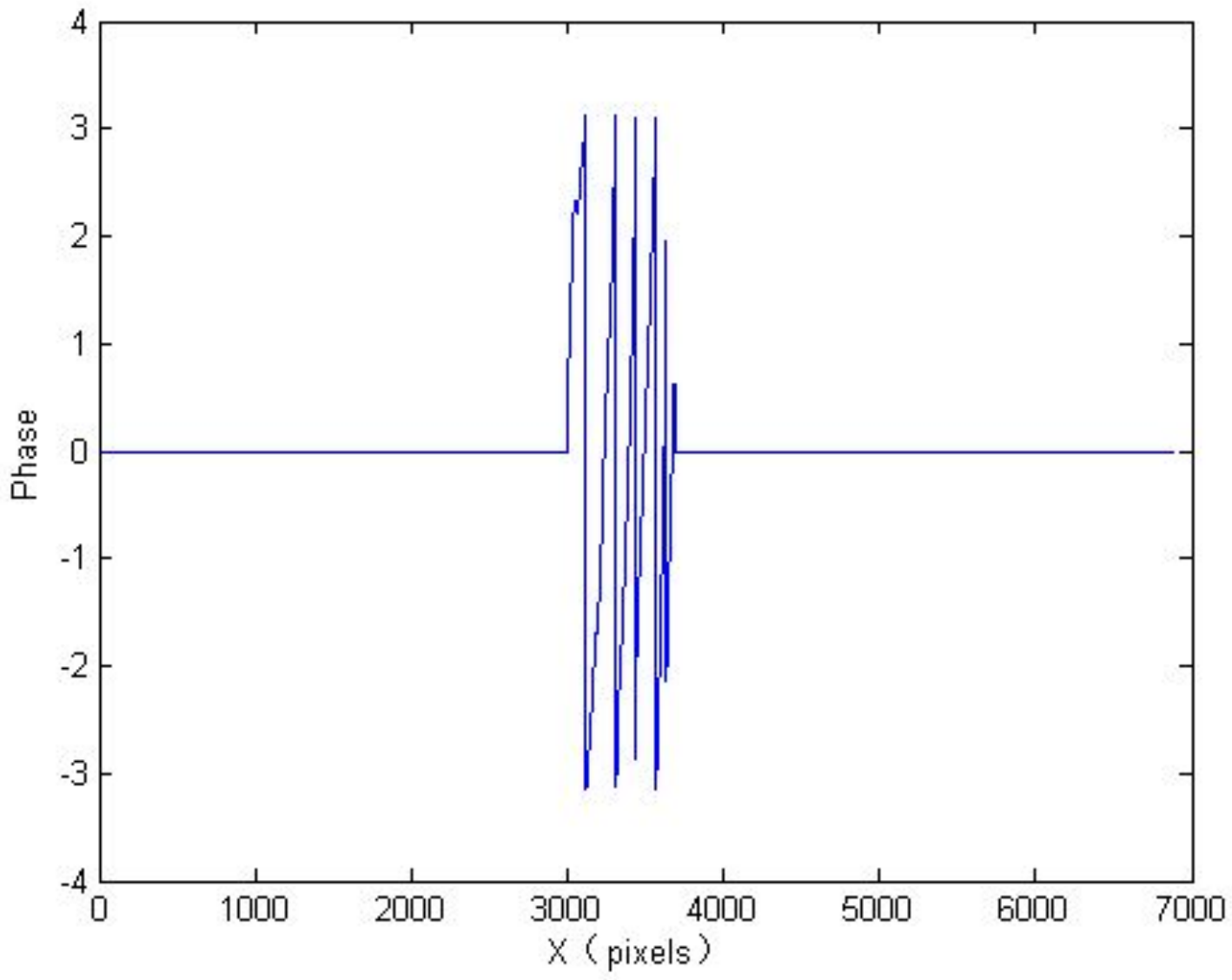

Fig. 10. The 300th row of Fig. 9 after zero padding

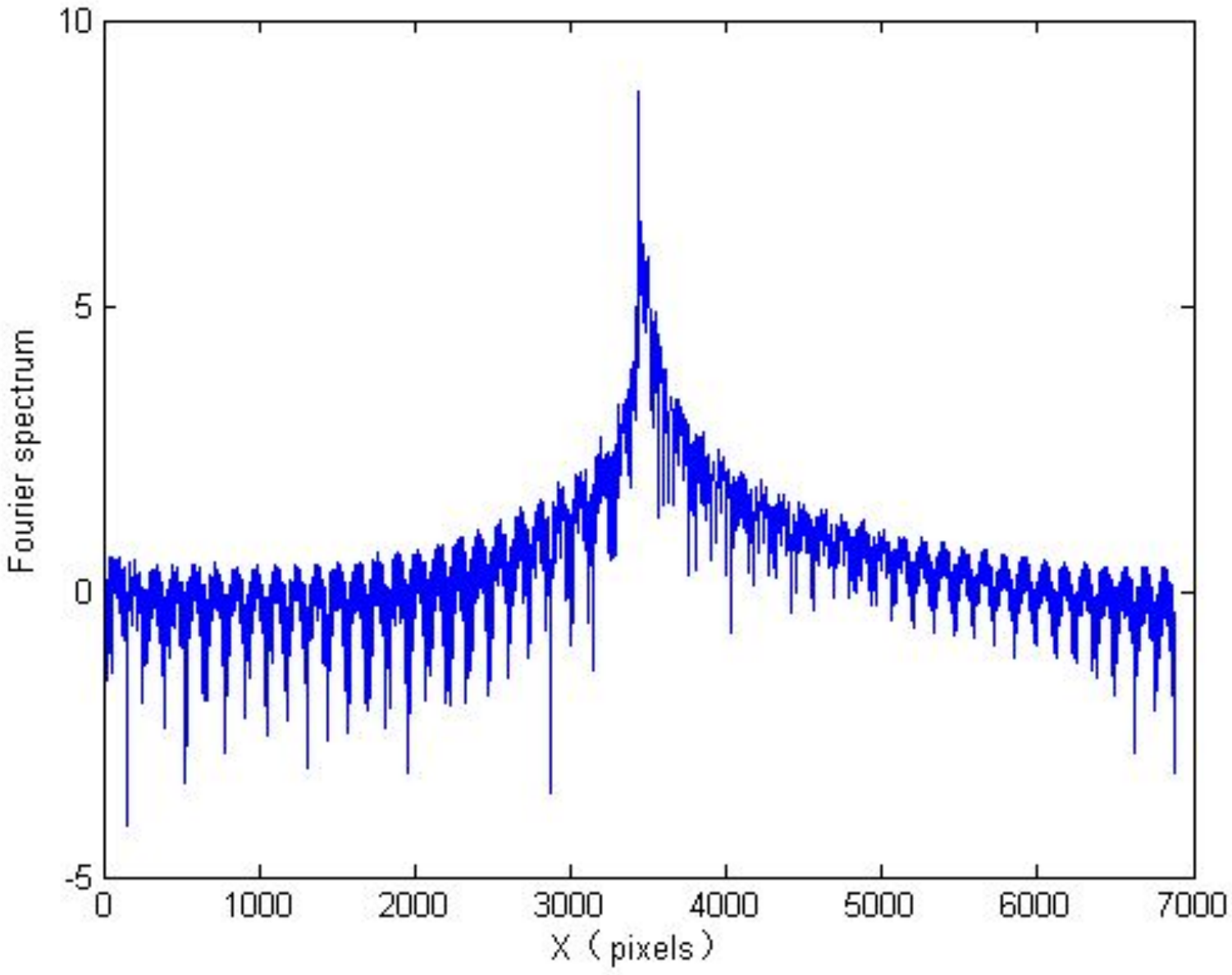

Fig.11. The spectrum of Fig. 10

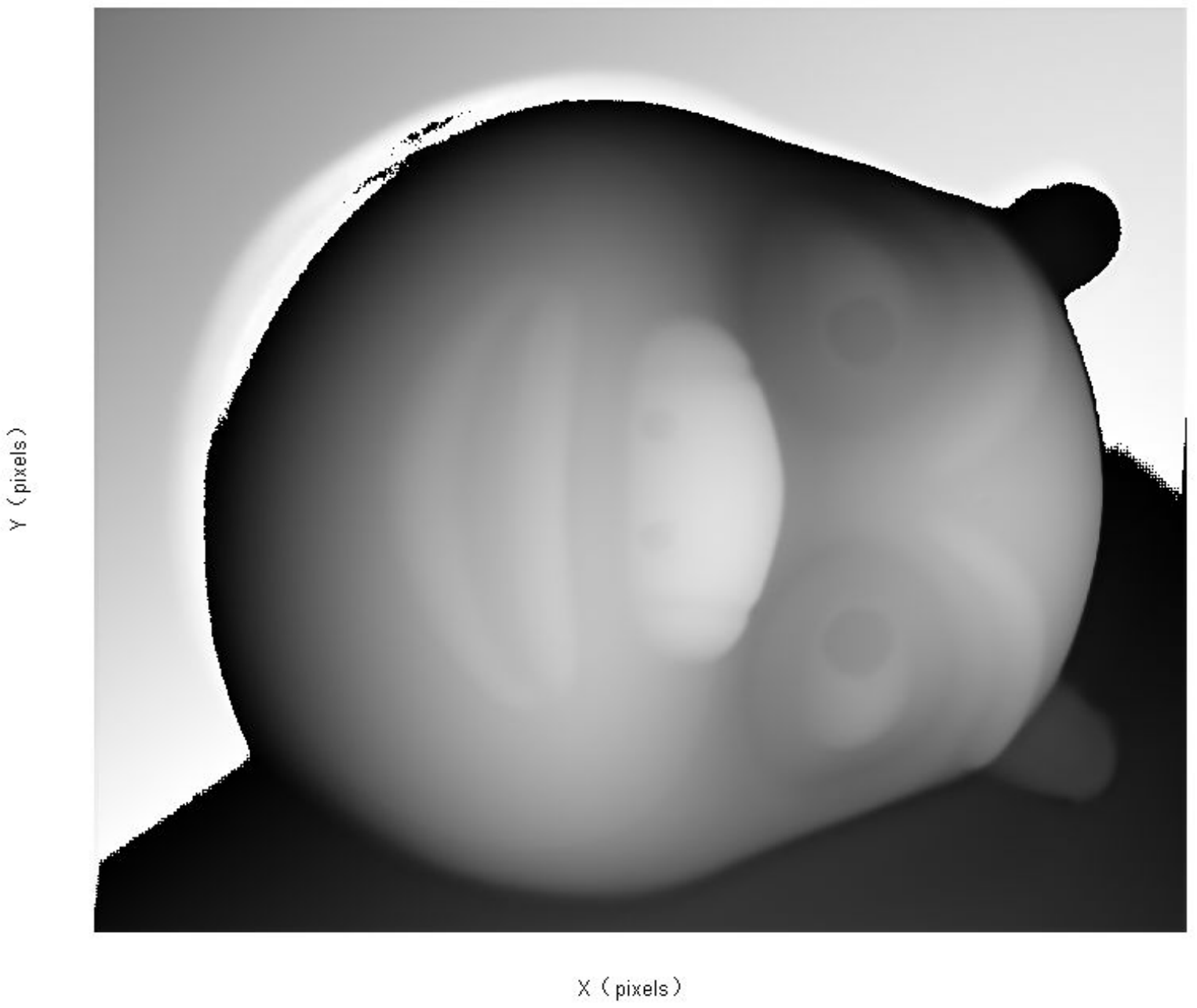

Fig.12. The phase map obtained by the proposed method

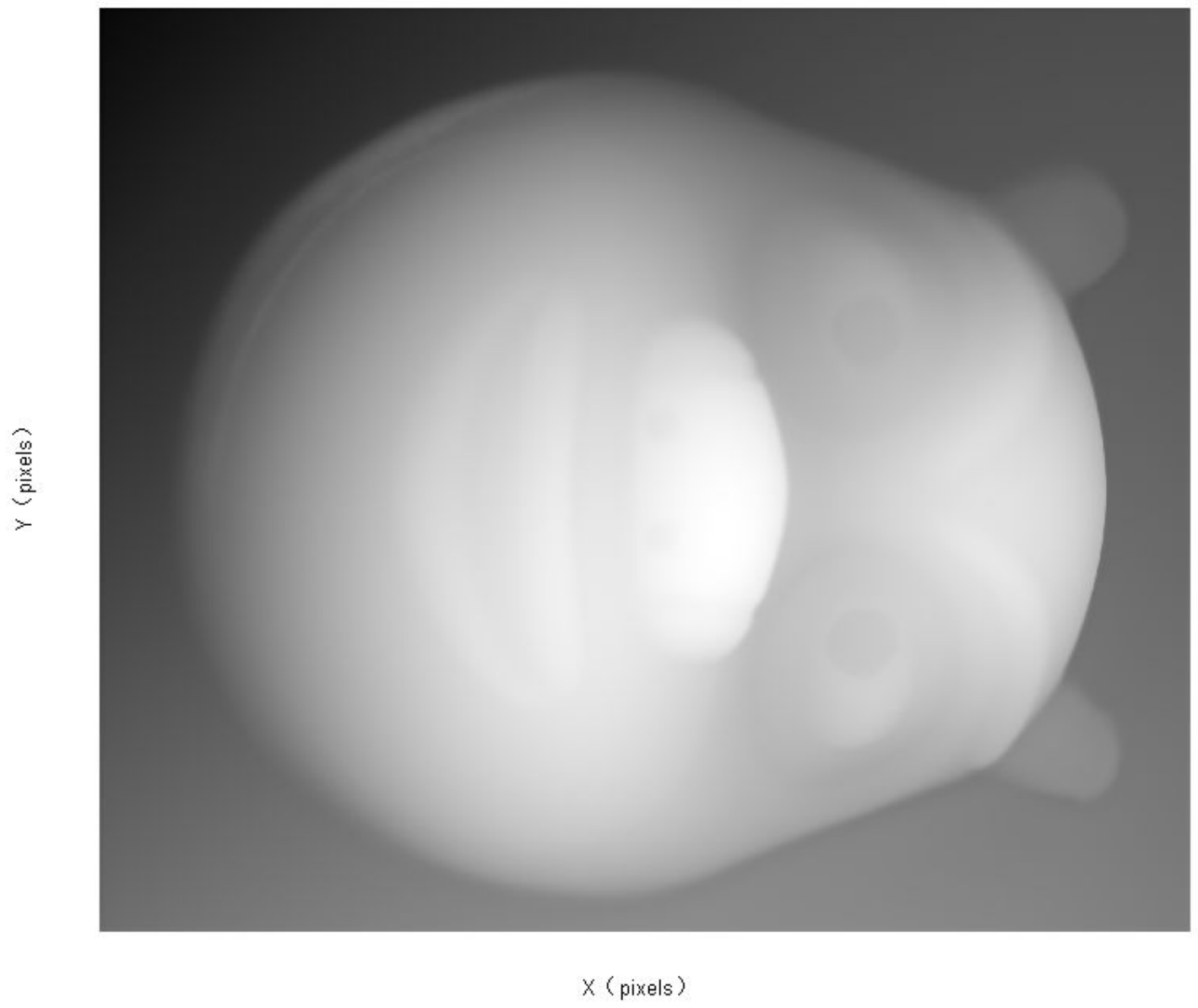

Fig. 13. The unwrapped phase map produced from Fig.12

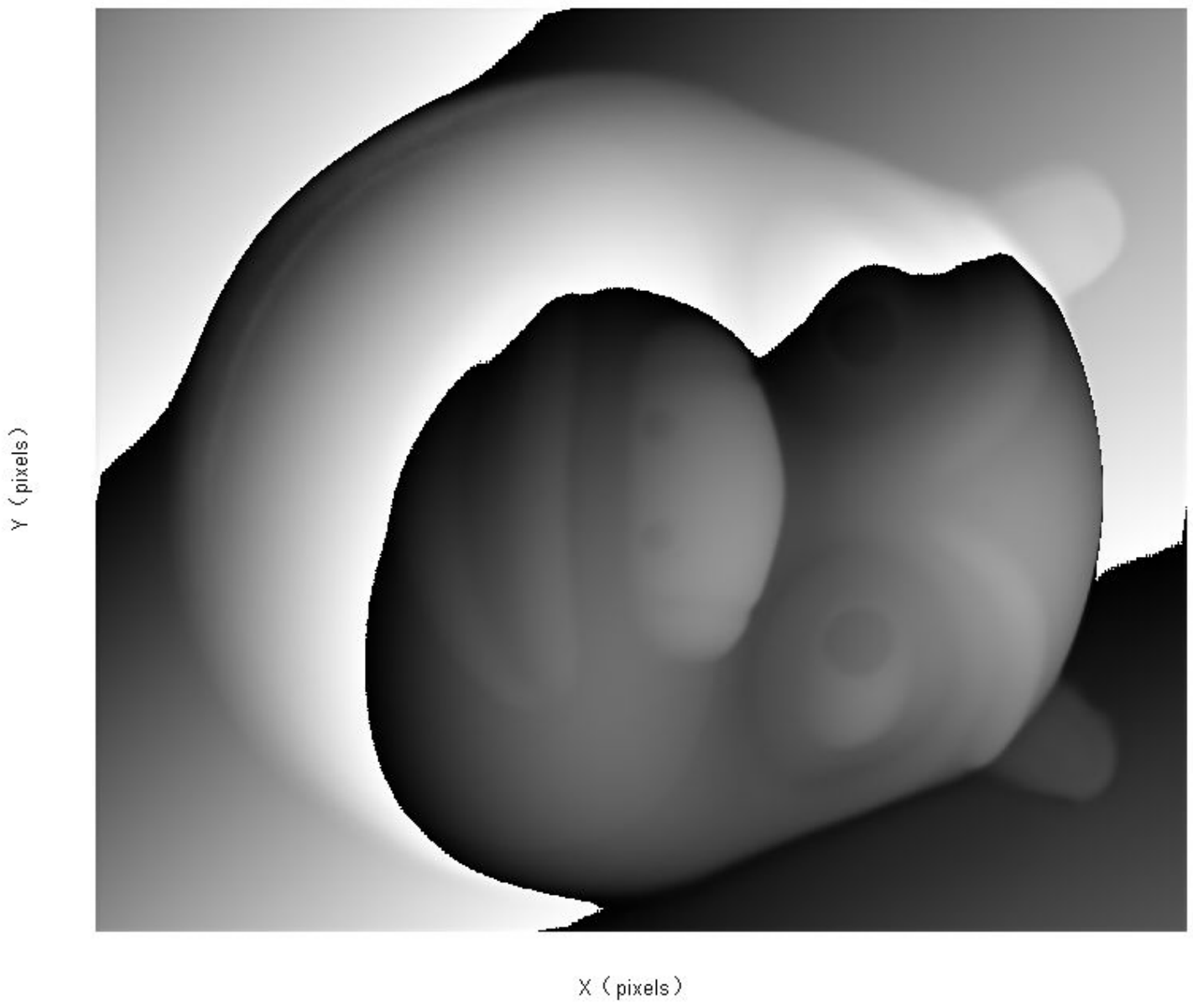

Fig. 14. The phase map obtained by Gdeisat's method

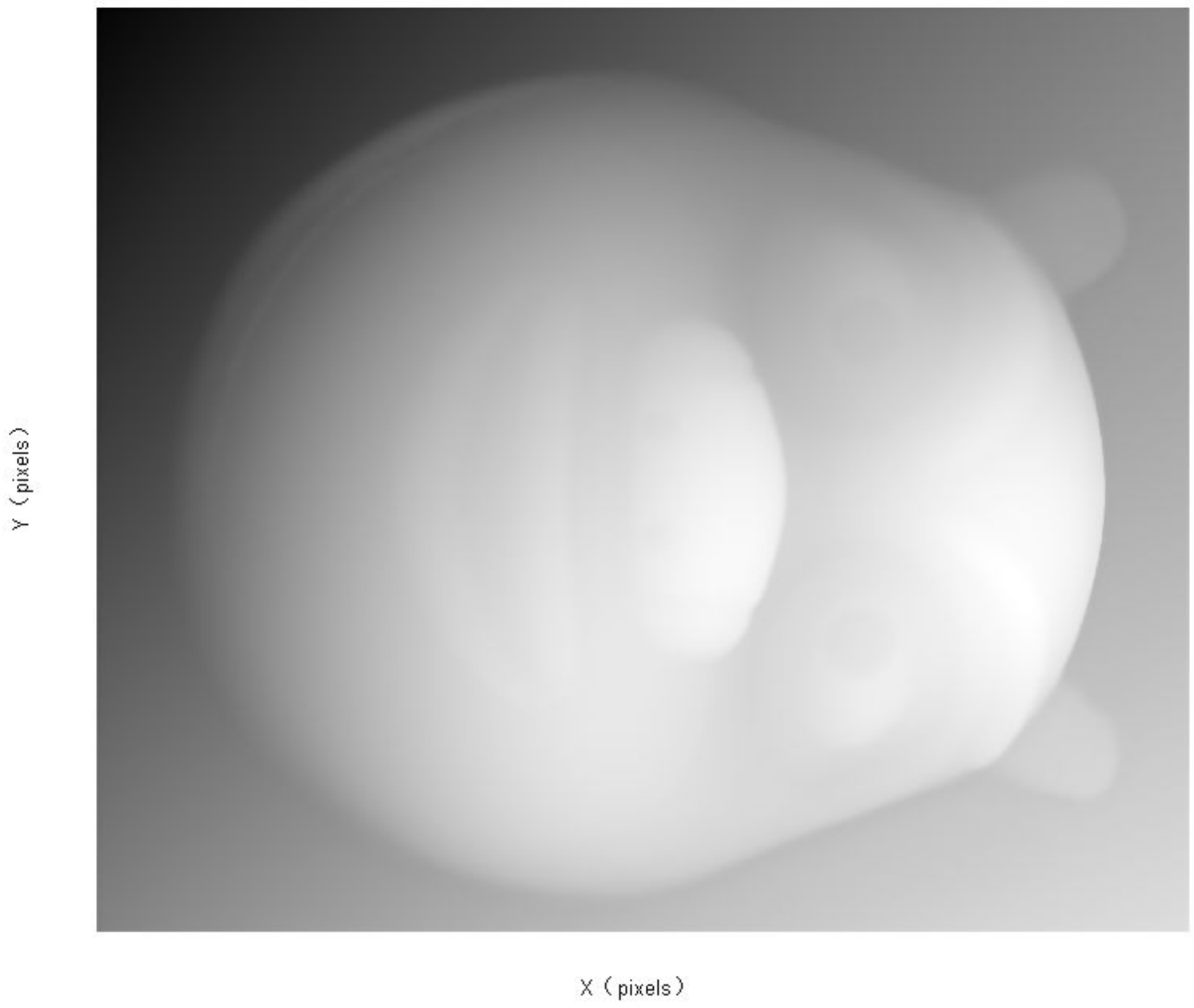

Fig. 15. The unwrapped phase map produced from Fig.14